\documentclass{elsart}

\usepackage{epsfig}

\usepackage{amssymb}

\begin{document}

\begin{frontmatter}


 \title{Coarse-graining polymers as soft colloids}


\author[Cam]{A.A. Louis\corauthref{cor1},}
\corauth[cor1]{Corresponding author. Tel.: +44-1223-336535;  fax:
+44-1223-336362}
\ead{aal20@cus.cam.ac.uk}
\author[Ams]{P.G. Bolhuis,}
\author[Cam]{R. Finken,}
\author[Cam]{V. Krakoviack,}
\author[Ams]{E.J. Meijer,}
\author[Cam]{J.P. Hansen}

\address[Cam]{Department of Chemistry, Lensfield Rd,
Cambridge CB2 1EW, UK}
\address[Ams]{Department of Chemical
Engineering, University of Amsterdam, Nieuwe Achtergracht 166, NL-1018
WV Amsterdam, Netherlands}

\begin{abstract}
We show how to coarse grain polymers in a good solvent as single
particles, interacting with density-independent or density-dependent
interactions.  These interactions can be between the centres of mass,
the mid-points or end-points of the polymers.  We also show how
to extend these methods to polymers in poor solvents and mixtures of 
polymers.  Treating polymers as soft colloids can greatly speed up the 
simulation of complex many-polymer systems, including  polymer-colloid mixtures.
\end{abstract}

\end{frontmatter}

\section{Introduction}
\label{intro}

Binary mixtures of colloidal particles and non-adsorbing polymers have
received renewed and growing attention recently, in part because they
exhibit complex and interesting structure, phase behaviour,
interfacial properties, and
rheology~\cite{Russ89,Lekk92,Meij94,Poon01,Loui01a,Liko01}, and in
part because they are excellent model systems for the study of large
length and time-scale separations in complex fluids.  Problems with
bridging length-scales are immediately apparent in even the simplest
models of colloid-polymer mixtures: while the mesoscopic colloidal
particles can be modeled as hard convex bodies, the polymers are
generally treated at the microscopic (Kuhn) segment level.  Thus, even
though the average size of the polymer coils may be of the same order
of magnitude as that of the colloids, the number of degrees of freedom
needed to model the former may be several orders of magnitude larger
than what is needed for the latter.  This naturally provokes the
question: Can the polymers also be modeled as single particles?  In
fact, this is exactly what was done by Asakura and Oosawa (AO) who, in
their classic work on colloid polymer mixtures\cite{Asak58}, modeled
the polymers as ideal particles with respect to each other, and as
hard-spheres with respect to the colloids.  This model is strictly
speaking only valid for non-interacting polymers, or for interacting
polymers in the dilute limit, while many interesting phenomena, such
as polymer induced phase separation, take place at finite
concentration of interacting polymers.  Our ultimate goal, therefore,
is to go well beyond the AO model and describe non-ideal polymers in a
good solvent up to semi-dilute concentrations.  We recently extended
the AO concept to take into account polymer-polymer interactions,
first by rather naively assuming a Gaussian repulsion between
polymers\cite{Loui99a} to account for the penetrable nature of polymer
coils, and then by carrying out a more sophisticated programme which
resulted in density-dependent\cite{Loui00,Bolh01,Bolh01a} and
density-independent\cite{Bolh01b} interactions between polymer coils.

The next sections will examine these effective potentials in more
detail. 

\section{Coarse-graining homogeneous polymer solutions}

Polymers made up $L$ segments are characterized by their radius of
gyration, $R_g \sim L^\nu$, where $\nu \approx 0.59$, i.e.\ polymers
are fractal objects.  Much of our understanding of polymer solutions
comes from the scaling arguments pioneered by de Gennes\cite{deGe79}.
These arguments suggest that the behaviour of a polymer solution
differs in the dilute regime, where the polymer coil density $\rho <
\rho^* = \frac43 \pi R_g^3$, from the semi-dilute regime, where
$\rho/\rho^* >> 1$.  Both these regimes presume that the actual
monomer concentration $c$ remains very low. Once $c$ becomes
appreciable, one enters the so-called melt regime which will not be
treated here, and for which different kinds of coarse-graining methods
are necessary.

A standard scaling argument suggest that the radius of gyration $R_g$
is the only relevant lengthscale for the dilute and semi-dilute
regimes of polymers in a good solvent\cite{deGe79}.  This immediately
implies that the second-virial coefficient should scale as $B_2 \sim
R_g^3$.  If we set $x = r/R_g$ then the second-virial coefficient is
proportional to
\begin{equation}\label{eq2.2}
B_2 \sim R_g^3 \int \left\{1 - \exp \left[ - \beta V(x)\right]\right\}
d{\bf x} \end{equation} where $\beta V(x)$ is the interaction between
two separate polymers, defined w.r.t.\ some yet to be specified
coordinate.  Since this must hold for all $R_g$, the interaction
$\beta V(r/R_g)$ should not depend on the length $L$ for sufficiently
long $L$ (scaling limit).  A more sophisticated version of this
argument was put forward by Grosberg, Khalatur, and
Khokhlov\cite{Gros82}, see also the review by Likos\cite{Liko01} for a
historical overview.

To coarse-grain each polymer as a single entity, one must still choose
an interaction centre, which may be the centre of mass (CM), the
mid-point, the end-points, or some average monomer.  For the mid-point
or end-point representation, $\beta V(x)$ should diverge at the origin,
since the actual segments of two different polymers cannot overlap.
For the CM, we expect a finite value of $\beta V(x=0)$, since it is possible
for two polymers to deform around each other in such a way that their
CM coincide without any mutually avoiding monomers overlapping.

\subsection{Density-independent polymer-polymer interactions}

In the description of atomic and molecular liquids and solids it is
common to replace the full quantum mechanical treatment of the
interactions by a simplified effective potential.  Well known examples
include the Lennard Jones pair potential and the Axilrod-Teller
three-body potential\cite{Hans86}.  Here we attempt a similar
coarse-graining for polymer solutions and choose the constituents to
be single polymers, with a specified interaction centre for each polymer.
Then, following for example \cite{Loui01a,Liko01} or more specifically
\cite{Bolh01a}, the coarse-grained Helmholtz free energy ${\sf F}$ of
a set of $N$ polymers with their interaction centres fixed at the
coordinates $\{{\bf r}_i\}$, in a volume $V$, can be written as the
following expansion:
\begin{eqnarray}\label{eqIII.1}
{\sf F}(N,V,\{{\bf r}_i\})& =& {\sf F}^{(0)}(N,V) + \sum_{i_1<i_2}^{N}
w^{(2)}({\bf r}_{i_1},{\bf r}_{i_2},) \\ \nonumber &+ & \sum^{N}_{i_1<i_2<i_3} w^{(3)}({\bf
r}_{i_1},{\bf r}_{i_2},{\bf r}_{i_3}) + \ldots + 
 w^{(N)}({\bf r}_{i_1},{\bf
r}_{i_2}\ldots {\bf r}_{i_N})
\end{eqnarray}
where the coordinates of the interaction centres of the polymers, $\{
{\bf r}_{i_1},{\bf r}_{i_2}\ldots {\bf r}_{i_n} \}$, are expressed in
units of $R_g$, the radius of gyration at zero density.  Each term in
this coarse-grained free energy includes an implicit statistical
average over all the internal monomeric degrees of freedom for a fixed
configuration $\{{\bf r}_i\}$.  ${\sf F}^{(0)}(N,V)$ is the so-called
volume term\cite{Liko01}, the contribution to the free energy that is
independent of the configuration $\{{\bf r}_i\}$, and includes the
internal free-energy of an isolated polymer.  For a homogeneous
solution, translational invariance implies that there is no one-body
term in the expansion.  Each subsequent term $w^{(n)}({\bf
r}_{i_1},{\bf r}_{i_2}\ldots {\bf r}_{i_n})$ is defined as the free
energy of $n$ polymers with their interaction  positions at
$\{ {\bf r}_{i_1},{\bf r}_{i_2}\ldots {\bf r}_{i_n} \}$, minus the
contributions of all lower order terms.  This procedure may in
principle be followed to derive higher and higher order interactions,
until, for a system with $N$ polymers, the $N$th term determines the
total coarse-grained free energy.  The thermodynamic free energy of
the
 polymer
solution finally follows from a statistical average over the
 interaction coordinates:
\begin{equation}\label{eqIII.1b} \beta F(N,V) = -\ln \sum_{\{{\bf
 r}_i\}} \exp \left[-\beta{\sf F}(N,V,\{{\bf r}_i\})\right].
\end{equation}
 But in practice, this approach is not often feasible because the
number of $n$-tuple coordinates and related complexity of each higher
order term increases rapidly with $n$, so that the series in
Eq.~(\ref{eqIII.1}) and the full average in Eq.~(\ref{eqIII.1b})
quickly become intractable.  Instead, one hopes to show that the
series converges fast enough  that only a few low order terms are
needed to obtain a desired accuracy.

\begin{figure}
\begin{center}
\epsfig{figure=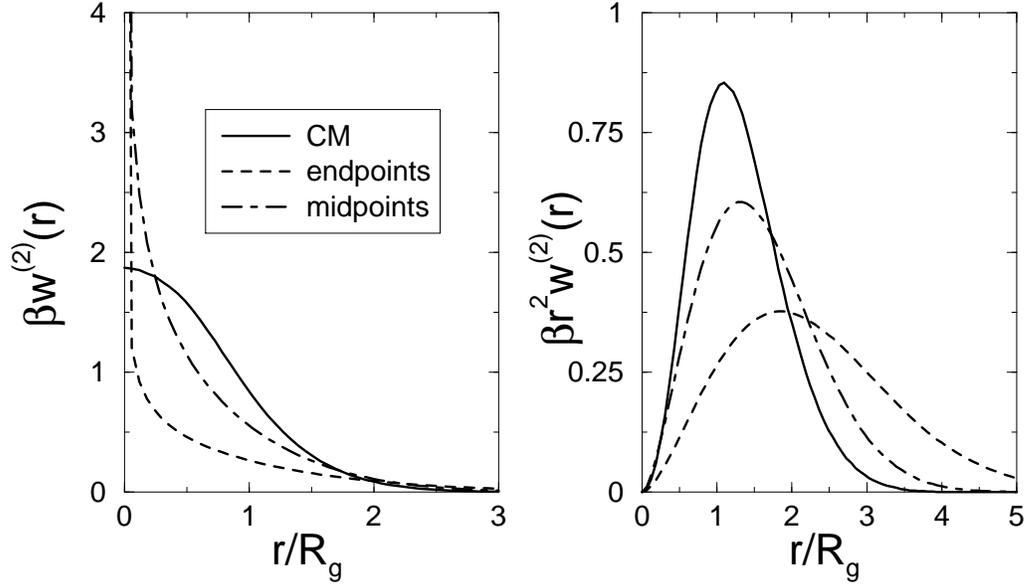,height=8cm}
\caption{\label{fig:vr-end-mid-CM} $w^{(2)}(r)$ and $r^2w^{(2)}(r)$
for the interaction between two isolated polymers in the CM,
end-point, and mid-point representations.  The end-point and mid-point
potentials diverge at the origin, but the CM representation gives a
finite value.  All three potentials result in the same second virial
coefficient $B_2$ defined in Eq.~(\protect\ref{eq2.2}).  }
\end{center}
\end{figure}

The first important term in the series expansion is the pair
interaction $w^{(2)}(r)$, which can be determined by calculating the
logarithm of the probability that two polymers have their interaction
centres a distance $r$ apart.  Details of our computer simulation
technique are described elsewhere\cite{Bolh01,Bolh01a,Rg}. In brief, by
simulating $L=500$ self-avoiding walk (SAW) polymers on a cubic
lattice, we determined $w^{(2)}(r)$ for three different interaction
centres: the end-points, mid-point, and CM of each polymer, as
depicted in Fig~\ref{fig:vr-end-mid-CM}.  The end and mid-point
representations diverge at the origin as $\lim_{r \rightarrow 0}
w^{(2)}(r) \sim \ln(r/R_g)$\cite{Liko01}, while the CM
representation has a finite value which we estimate to be $w^{(2)}(0)
= 1.80 \pm 0.05$ in the scaling limit $L \rightarrow
\infty$\cite{Bolh01}.  By plotting $r^2 v(r)$ we see that the CM
representation has the shortest range, which is one reason why it is
easier to use than the other two representations.

In a similar fashion, the higher order interactions can be calculated
from higher order probability distributions\cite{Loui01a,Bolh01a}.  We
calculated the relative strength of the many-body terms up to fifth
order by computer simulations, and to arbitrary order by a scaling
theory\cite{Bolh01a}.  The simulations and the scaling theory agree
well, and suggest that at full overlap the $nth$ order many-body term
alternates in sign as $(-1)^n$, and decreases (slowly) in absolute
magnitude with increasing $n$.  But, as mentioned before, a
description based on three and higher order interactions rapidly
becomes too unwieldy to use.  Instead, we show in the next section how
to derive a pair potential approach which includes these higher order
($n>2$) interactions in an average way.

\subsection{Density-dependent polymer-polymer interactions}

An alternative coarse-graining approach to the many-body expansion of
the previous section is to find pair potentials which reproduce known
structural information.  We are aided in this by a theorem which
states that at a given density $\rho$, there is a one-to-one mapping
between the pair distribution function $g(r)$ and a unique pair
potential $v(r;\rho)$ that will exactly reproduce the correct pair
correlations\cite{Hend74}.  We generate the pair correlations with
simulations of SAW polymers, and at each density $\rho$, use the
Ornstein-Zernike equations\cite{Hans86}, coupled with the
hypernetted-chain closure (HNC)\cite{Hans86}, to invert the CM $g(r)$
and find  $v(r;\rho)$.  While the HNC closure is generally not
accurate enough for inversions in simple liquids, it is nearly exact
for the soft potentials we are investigating
here\cite{Loui01a,Liko01,Loui00a}.  Nevertheless, there are a number
of subtleties, both in the simulations and in the inversions, which must
be carefully examined\cite{Bolh01,Bolh01b}.

\begin{figure}
\begin{center}
\epsfig{figure=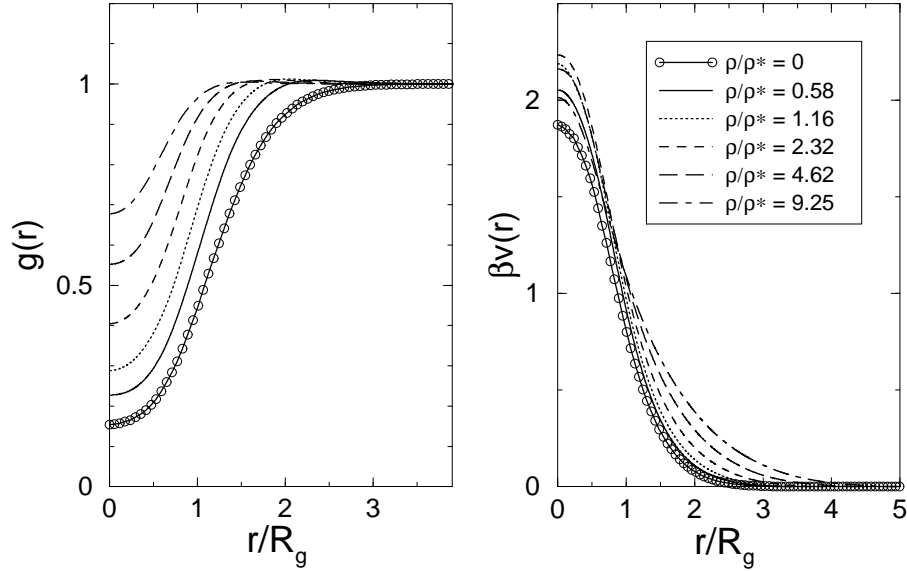,width=12cm}
\caption{\label{fig:veffL500} The effective polymer pair potentials
$v(r;\rho)$, derived at different densities from an HNC inversion of
the CM pair distribution functions $g(r)$ of $L=500$ SAW polymer
coils. (from Ref\protect\cite{Bolh01a})}
\end{center}
\end{figure}

Density-dependent effective potentials $v(r;\rho)$, inverted from the
$g(r)$ produced by $L=500$ SAW simulations are shown in
Fig.~(\ref{fig:veffL500}).  The potential changes with increasing
density, but approximately retains the shape found at $\rho=0$.  We
have recently shown that these potentials can be very accurately
parameterized for $\rho/\rho^* < 2$ by sums of three Gaussians with
density-dependent coefficients\cite{Bolh01b}.

Within the HNC approximation, the density dependence of an effective
pair potential that reproduces the true $g(r)$ is given to lowest
order in $\rho$ and the $w^{(n)}(\{{\bf r}_i\})$ by\cite{vdHo99}:
\begin{equation}
\label{eq:axilrod}
v(r_{12};\rho)  =  w^{(2)}(r_{12}) - \rho \int \left(
e^{-\beta w^{(3)}(r_{12},r_{13},r_{23})} -1 \right) 
g_2(r_{13};\rho) g_2(r_{23};\rho) d{\bf r}_3,
\end{equation}
We found that this expression describes the density dependence quite
well for $\rho/\rho^*<1$, and even works qualitatively for higher
densities, where we expect higher order $\rho$ and $w^{(n)}(\{{\bf
r}_i\})$ effects to become significant\cite{Bolh01a}.  This demonstrates the
connection between the density-independent and density-dependent
approaches, showing explicitly that the density dependence in the
effective pair-potentials $v(r;\rho)$ arises from the many-body
interactions.

\begin{figure}
\begin{center}
\epsfig{figure=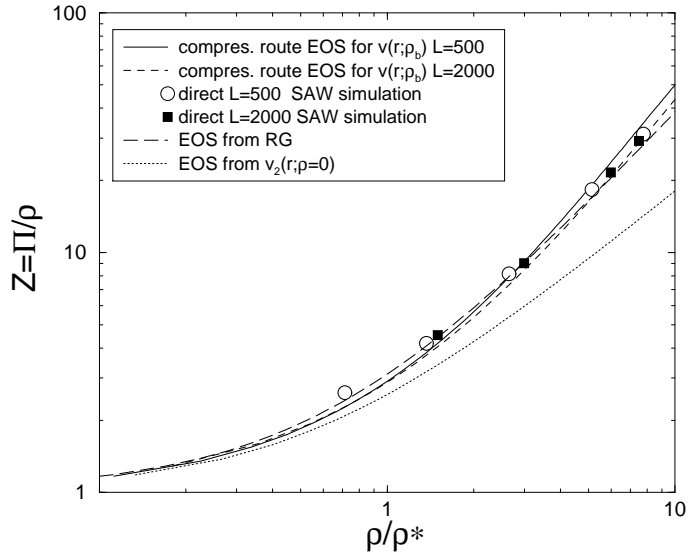,width=9cm}
\caption{\label{fig:Z-linear} EOS for polymers in good solvent.
Direct results and the EOS arising from the effective potentials
through the compressibility route are compared for $L=500$ and
$L=2000$ SAW polymers.  }
\end{center}
\end{figure}

One advantage of the structure-based route to the potentials is that
one can use the compressibility equation\cite{Hans86} to derive the
equation of state (EOS) from the pair correlations.  We have done this
for both $L=500$ and $L=2000$ SAW simulations.  We directly measured
the EOS and compared this to the EOS derived from the effective
potentials through the compressibility equation\cite{Bolh01,Bolh01b}.
The two routes are compared in Fig.~\ref{fig:Z-linear}, where the
agreement is shown to be excellent.

 All three effective potentials shown in Fig~\ref{fig:vr-end-mid-CM}
result in ``mean-field fluids''\cite{Loui01a,Liko01,Loui00a}, so named
because the EOS takes on the mean-field form $\beta \Pi/\rho \sim 1 +
\rho \hat{v}(k=0)$ at high enough densities. Here $\hat{v}(k)$ is the
Fourier transform (FT) of the potential. This implies that if we only
use the $\rho=0$ potential, then the EOS at higher density would scale
as $\beta \Pi /\rho \sim \rho$ instead of the the correct $\rho\sim
\rho^{1.3}$ scaling found for the semi-dilute regime\cite{deGe79}.  It
is therefore the many-body interactions, expressed through the density
dependence of $v(r;\rho)$, which cause the EOS to be super-linear.

 We add a caveat here about the route to thermodynamics with density
dependent potentials.  The potentials derived here can be used to
derive the correct thermodynamics through the compressibility route.
Different (but related) density-dependent potentials, which do not
reproduce the correct structure, would be needed to derive the correct
thermodynamics through the virial route\cite{vdHo99}.  In other words,
there is no unique density-dependent pair potential: when specifying
such a potential, one must also specify which route to thermodynamics
should be used\cite{Loui01c}.

\section{Polymers near walls and spheres}

\begin{figure}
\begin{center}
\epsfig{figure=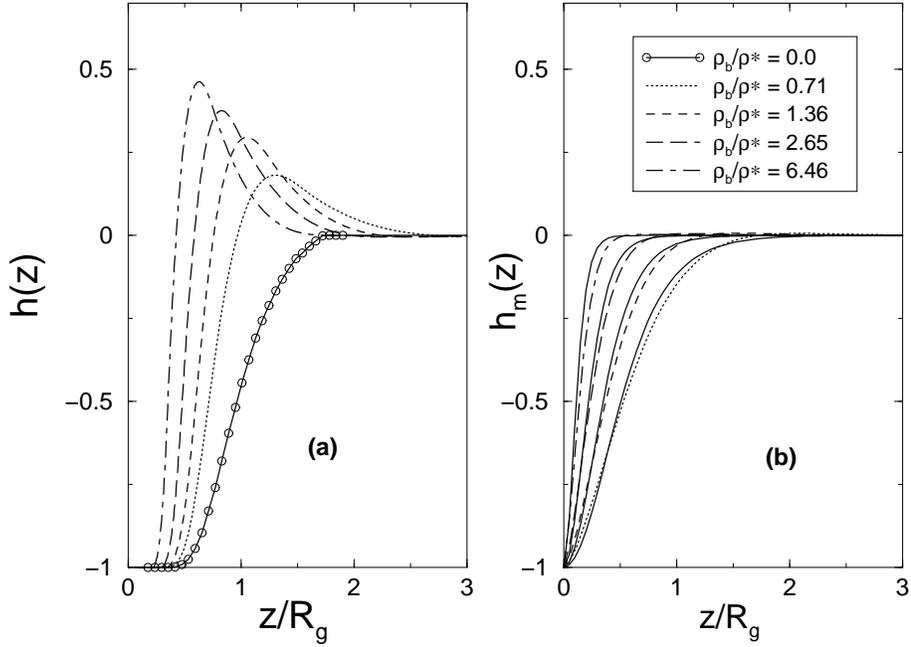,width=12cm}
\caption{\label{fig:hz} {\bf (a)} The wall-polymer CM profile $h(z) =
\rho(z)/\rho -1$ for $L=500$ SAW polymers at different bulk
concentrations.  {\bf (b)} The wall-polymer monomer profile $h_m(z)$
for the same bulk concentrations.  Both representations result, by
definition, in the same relative adsorptions $\Gamma/\rho$.  The
straight lines in {\bf (b)} are a fit to the simple form $h_m(z) =
\tanh^2(-z\rho/\Gamma(\rho)) -1$\protect\cite{deGe79}}
\end{center}
\end{figure}

Polymers form a depletion layer near a hard non-adsorbing wall because
the number of possible conformations are restricted there. This is
illustrated in Fig~\ref{fig:hz} for the CM and for a monomer
representation near a planar wall.  We have used an inversion method
similar to that used in the previous sections to derive effective
wall-polymer potentials for interacting polymers near walls and
spheres\cite{Loui00,Bolh01,Bolh01b}.  These potentials are constrained
to give the correct density profile $\rho(r)$ which in turn determines
the adsorption $\Gamma$.  One can show, for example, that
the surface tension is completely determined if one knows the EOS and
$\Gamma$ as a function of $\rho$\cite{Mao97}.  Our effective polymer-polymer
pair potentials correctly determine the EOS, while the wall-polymer
potentials correctly determine $\Gamma$, implying that our formulation
will reproduce the correct surface thermodynamics.

\begin{figure}
\begin{center}
\epsfig{figure=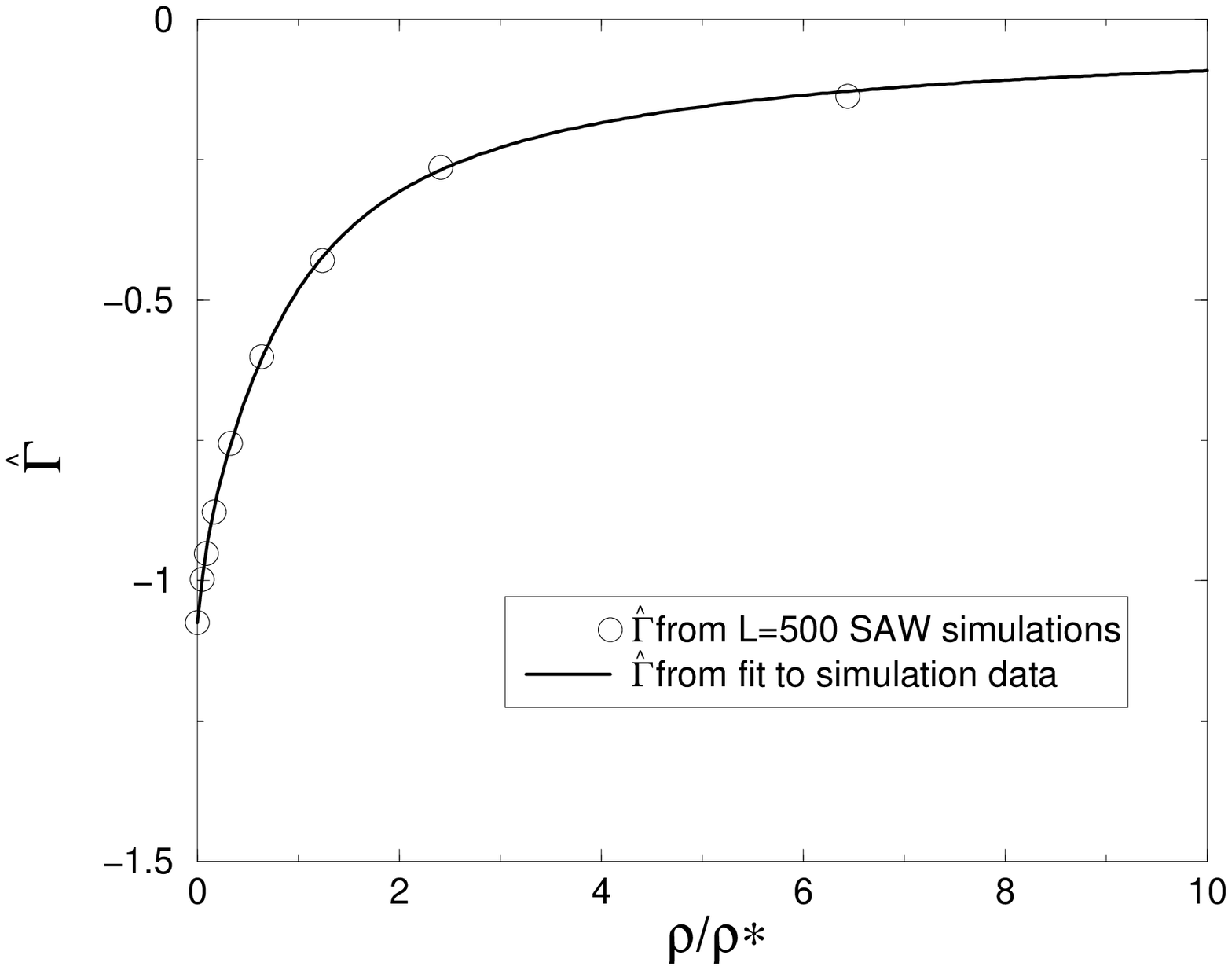,width=8cm}
\caption{\label{fig:Gamma-wall-fit} Relative adsorption $\Gamma/\rho$,
in units of $R_g$ of $L=500$ SAW polymers near a single hard
wall. Circles are direct simulations and the line denotes the simple
fit with the correct scaling behavior, given by $
\hat{\Gamma}(\rho) =- 1.074 R_g\left( 1 + 7.63\frac{\rho}{\rho^*} +
14.56(\frac{\rho}{\rho^*})^3\right)^{-(0.2565)}$
}
\end{center}
\end{figure}

We show this adsorption $\Gamma/\rho$ in
Fig.~\ref{fig:Gamma-wall-fit}, together with a simple fit constrained
to give the correct scaling $\Gamma/\rho \approx \xi(\rho) \sim \rho^{-0.77}$
in the semi-dilute regime.  Note that the largest relative
change in the adsorption is actually in the dilute regime, suggesting
that even there descriptions based on the low-density or
non-interacting polymer limit rapidly become inadequate.

\section{Connection with scaling theory}

Most successful theories of polymers start from a monomer based
description and use scaling or RG approaches to derive properties of
polymer solutions\cite{deGe79}.  How does our CM based description
compare with these scaling approaches?  For example, in the semi-dilute
regime, scaling theories predict that the important length-scale is
the correlation length $\xi(\rho)$, which decreases with increasing
density as $\xi(\rho) \sim \rho^{-0.77}$.  It is not a-priori clear
how this lengthscale enters into the $g(r)$ or the $v(r)$ in our
description of homogeneous polymer solutions. The EOS scales as $\beta
\Pi/\rho \sim \xi^{-3}/\rho$ in the semi-dilute regime, a behaviour
which is reproduced by our description through the compressibility
equation.  Since these potentials result in ``mean-field fluids'', this
suggests that  $\int dr  r^2 v(r;\rho) \sim \xi^{-3}/\rho^2$ in the
semi-dilute regime.  For inhomogeneous systems in the semi-dilute
regime, $\xi$ enters more directly through the density profiles shown
in Fig~\ref{fig:hz}, but again the direct connection to the potentials
is more opaque.  So the exact connection with the scaling theory still
remains to be worked out.  We expect our approach to be most robust in
the dilute regime and into the crossover region of the semi-dilute
regime. Luckily this is also where much of the interesting physics of
the colloid-polymer systems  lies.  How well our
``soft colloids'' approach will work deep into the semi-dilute regime
still remains to be established.

\section{Extensions to  poor solvents and mixtures}
The considerations in the previous sections focused on equal length
polymers in a good solvent, where the temperature plays no role.
However, the techniques used for polymers in a good solvent should
still apply to other types of polymers solutions.

\subsection{Poor solvents}
\begin{figure}
\begin{center}
\epsfig{figure=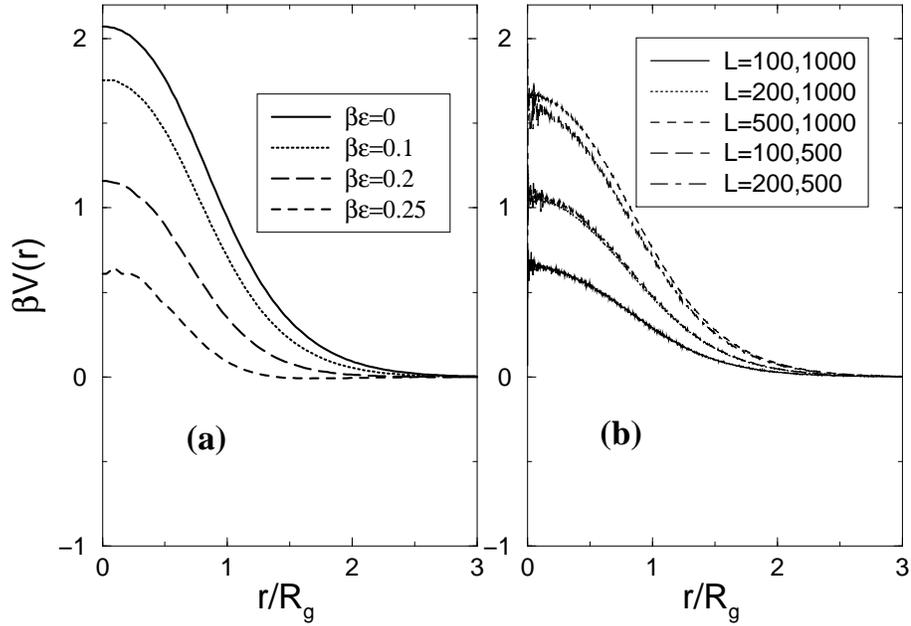,width=12cm}
\caption{\label{fig:poorsolvent}{\bf (a)} CM-CM interaction for
polymers in a poor solvent, modeled by $L=100$ SAW polymers on a
cubic lattice with nearest neighbour attractions of strength $-\beta
\epsilon$. {\bf (b)}CM-CM interaction for mixtures of different length
SAW polymers. }
\end{center}
\end{figure}

We first examine briefly what happens for polymers in a poor solvent,
using as a model SAW polymers with a nearest neighbor attraction of
strength $-\beta \epsilon$.  It is known that as the temperature
decreases, there is a temperature $T_{col}$ below which the polymer
collapses into a compact globule and loses its fractal
nature\cite{deGe79}.  The effective potentials will then be
fundamentally different of course.  But as long as we stay above this
temperature, we expect that the interaction should become less strong
with decreasing temperature, as shown in Fig.~\ref{fig:poorsolvent}.

\subsection{Mixtures}

Renormalization group (RG) calculations for the interaction between
the CM of two polymers of differing lengths $R_{g1}$ and $R_{g2}$
suggest that the interaction strength at full overlap should weaken
with increasing size asymmetry, and that the interaction range should
approximately scale as $R_{12} = \frac12 \sqrt{R_{g1}^2 +
R_{g2}^2}$\cite{Krug89}.  We confirm this behaviour for
simulations of a number of different length SAW polymers at $\rho=0$
in Fig.~\ref{fig:poorsolvent}.

\subsection{Phase separation?}

An interesting prediction is that binary mixtures of Gaussian core
particles interacting via the pair potentials $V_{\alpha \beta}(r) =
\epsilon_{\alpha \beta} \exp \{ -(r/R_{\alpha \beta})^2 \}$ will
phase-separate over broad ranges of the coupling constant ratio
$|\epsilon_{12}|/\sqrt{\epsilon_{11} \epsilon_{22}}$.
Semi-quantitative correct phase-diagrams follow already from an
analytic mean-field calculation\cite{Fink01}, and suggest that 
polymers in a good solvent will not phase separate at low densities.

\subsection{Relationship with PRISM}
Each time the present coarse-graining methods are applied to a new type of
polymer solution we need a new set of computer simulations at the
monomer level for the parameters of interest. It would be very helpful
to find other, semi-analytic, ways of providing input information that
are faster and more flexible.  One candidate would be
PRISM\cite{Schw97}, an integral equation method which has been applied
to a wide variety of polymeric systems.  This requires a way of
deriving CM-CM correlations from the monomer based correlations
provided by PRISM. As a first step in this direction, we have derived
an approximate relationship which is much more accurate than earlier,
heuristic approaches\cite{Krak01}. This could form the basis for using
PRISM or other monomer based methods as input to our ``polymers as
soft colloids'' approach.

\section{Conclusions}

In summary then, we have shown how to coarse-grain polymers as single
``soft colloids'', with just three degrees of freedom each,
interacting via a density-independent pair, triplet, and higher body
potentials. These, however, become rapidly intractable.  In parallel,
we also derived density-dependent pair potentials which include, in an
average way, the effect of the higher $n$-body interactions.  These
effective pair potentials exactly reproduce the two-body correlations
of the underlying polymer solution, and, through the compressibility
equation, they reproduce the EOS as well.  In a similar way the
effective one-body potentials which exactly reproduce the one-body
density profiles near walls and spheres can also be derived.  Because
these reproduce the correct adsorptions, the thermodynamics of a
polymer solution near a non-adsorbing surfaces are also correctly
reproduced by our formulation.

We also showed how to extend these methods to derive effective
potentials for polymers in a poor solvent and for mixtures of
different length polymers.  We sketched some ways in which other
monomer based methods such as PRISM could be used as the source of
input to derive our potentials.

At this point one might ask what has been gained, since at each point
direct computer simulations were needed as input to derive the
potentials.  This question brings us back to the aim stated at the
outset: to describe mixtures of many polymers and many colloids.  Here
our coarse-graining of polymers as soft colloids does result in
important simplifications.  For example, we have performed such
simulations for mixtures of spheres of radius $R_c$ and polymers with
sizes ranging from $R_g/R_c \approx 0.3$ to $R_g/R_c \approx 1$, and
determined the polymer induced phase-separation of the
colloids\cite{Bolh01d}.  The effective polymer density $\rho/\rho^*$
at the critical point increases with increasing polymer size, but even
for the largest polymers it is still in the regime $\rho/\rho^* \leq
1$, where we expect our formulation to work best.  Simulations with a
full polymer model would be about two orders of magnitude slower.
Without coarse-graining the polymers as soft colloids, such a
simulation would have been virtually impossible to perform.

\ack AAL acknowledges support from the Isaac Newton Trust, Cambridge,
PB and VK acknowledge support from the EPSRC under grant number
GR$/$M88839, RF acknowledges support from the Oppenheim Trust, EJM
acknowledges support from the Royal Netherlands Academy of Arts and
Sciences. EJM acknowledges support from the Stichting Nationale
Computerfaciliteiten (NCF) and the Nederlandse Organisatie voor
Wetenschappelijk Onderzoek (NWO) for the use of supercomputer
facilities.

\begin{figure}
\begin{center}
\epsfig{figure=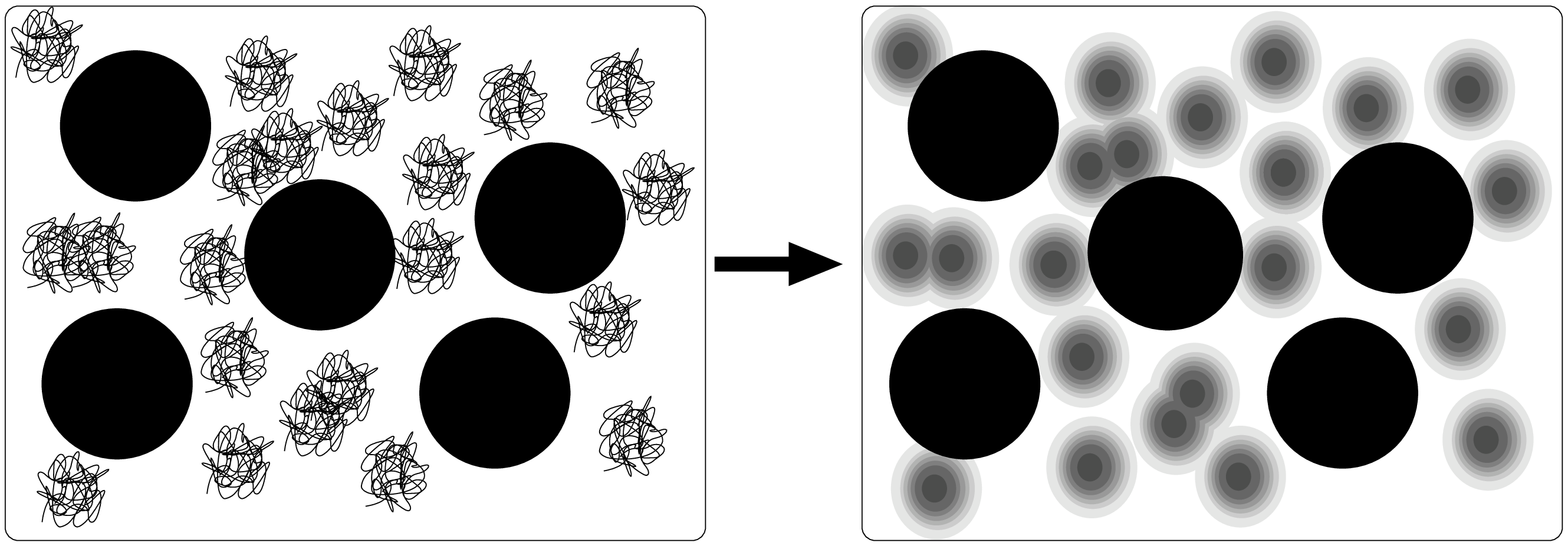,width=12cm,angle=-90}
\caption{\label{fig:col-pol} Schematic picture of our coarse-graining
scheme: The polymers in a polymer-colloid mixture are treated as
+single  entities, on the same footing as the colloidal
particles. }
\end{center}
\end{figure}

\begin{thebibliography}{00}
\bibitem{Rg}In our earlier work\protect\cite{Loui00,Bolh01,Bolh01a}
we used $R_g = 16.5$ instead of the correct $R_g=16.83$ value for the 
$L=500$ SAW polymers.  This means a slight adjustment in the values 
of $R_g$ and in $\rho/\rho*$ quoted in our papers.  For example, 
the values quoted for $\rho/\rho*$ should increase by a factor $1.06$.
\end{thebibliography}
\end{document}